# Pnictogens Allotropy and Phase Transformation during van der Waals Growth

Matthieu Fortin-Deschênes[1], Hannes Zschiesche[2], Tevfik O. Menteş[3], Andrea Locatelli[3], Robert M. Jacobberger[4], Francesca Genuzio[3], Maureen J. Lagos[2], Deepnarayan Biswas[5], Chris Jozwiak[6], Jill A. Miwa[5], Søren Ulstrup[5], Aaron Bostwick[6], Eli Rotenberg[6], Michael S. Arnold[4], Gianluigi A. Botton[2], Oussama Moutanabbir[1]

[1] Department of Engineering Physics, École Polytechnique de Montréal, C. P. 6079, Succursale Centre-Ville, Montréal, Québec, H3C 3A7, Canada

[2] Department of Materials Science and Engineering, McMaster University, Hamilton, Ontario, L8S 4L8, Canada

[3] Elettra Sincrotrone Trieste S.C.p.A, S.S. 14 – km 163, 5 in AREA Science Park, 34149 Basovizza, Trieste, Italy

[4] Department of Materials Science and Engineering, University of Wisconsin-Madison, Madison, Wisconsin 53706, USA

[5] Department of Physics and Astronomy, Interdisciplinary Nanoscience Center, Aarhus University, 8000 Aarhus C, Denmark

[6] Advanced Light Source, Lawrence Berkeley National Laboratory, Berkeley, CA 94720, USA

Pnictogens have multiple allotropic forms resulting from their ns2 np3 valence electronic configuration, making them the only elemental materials to crystallize in layered van der Waals (vdW) and quasi-vdW structures throughout the group[1-5]. Light group VA elements are found in the layered orthorhombic A17 phase such as black phosphorus[1,3], and can transition to the layered rhombohedral A7 phase at high pressure[6,7]. On the other hand, bulk heavier elements are only stable in the A7 phase[2,4]. Herein, we demonstrate that these two phases not only co-exist during the vdW growth of antimony on weakly interacting surfaces, but also undertake a spontaneous transformation from the A17 phase to the thermodynamically stable A7 phase. This metastability of the A17 phase is revealed by real-time studies unraveling its thickness-driven transition to the A7 phase and the concomitant evolution of its electronic properties. At a critical thickness of ~4 nm, A17 antimony undergoes a diffusionless shuffle transition from AB to AA stacked α-antimonene followed by a gradual relaxation to the A7 bulk-like phase. Furthermore, the electronic structure of this intermediate phase is found to be determined by surface self-passivation and the associated competition between A7- and A17-like bonding in the bulk. These results highlight the critical role of the atomic structure and interfacial interactions in shaping the stability and electronic characteristics of vdW layered materials, thus enabling a new degree of freedom to engineer their properties using scalable processes.

The thermodynamic behavior of vdW and quasi-vdW materials is intrinsically linked to their unique layered structure as their stability is insured by their strong in-plane bonds. For instance, the sp$^2$ bonding in graphite and BN stabilizes the layered phases at the expense of the diamond phase[8,9]. On the other hand, the low density associated with vdW interlayer interactions makes layered phases increasingly unstable at high pressure. Moreover, since these phases usually have lower symmetry and a more complex atomic structure, they tend to be stabilized by entropic contributions at high temperature. Generally, the thermodynamic properties of a material are greatly altered at nanoscale dimensions because of the increasing importance of surface energy contributions[10-13]. However, due to their self-passivated nature and low surface energies, shrinking the dimensions of vdW layered materials does not significantly impact their stability. Therefore, it is reasonable to anticipate that the growth of polymorphic materials on weakly interacting surfaces would favor the nucleation of layered phases that are metastable at larger thicknesses, but persist due to the important phase transformation energy barriers induced by the strong directional in-plane bonds. Nonetheless, the unique electronic and atomic structure of group VA elements can facilitate these early metastable to stable phase transformations.

There are two classes of elemental materials that crystallize in layered structures: group IVA and group VA solids[1-5]. For group IVA materials, only sp$^2$ hybridized C has a layered structure (graphite). Other group IVA elements prefer the diamond cubic structure with sp$^3$ hybridization. Group VA elements also tend to favor sp$^3$ hybridization (except for sp hybridized nitrogen). However, with their ns2 np3 valence configuration, they form quasi-vdW layered structures with three covalent in-plane bonds and weaker interlayer bonds. This electronic configuration leads to two stable layered crystal structures: orthorhombic A17 and rhombohedral A7 (Fig. 1a). Phosphorus is most stable in the A17 phase (black phosphorus) in ambient conditions[1]. At higher pressures, it transitions to the A7 phase[6]. The transformation is hypothesized to occur by a shuffle and shear mechanism[14,15]. However, the exact atomic pathways of this mechanism remain in the realm of speculations due to the lack of direct experimental evidence. Arsenic (As) is more stable in the A7 phase[2], but can also crystallize in the A17 phase[16]. The A17 to A7 phase transformation was observed under extreme conditions involving high pressure[7]. With the increase in interlayer interactions with atomic number, the A7 phase becomes more stable for the heavier elements of antimony (Sb) and bismuth (Bi) due to its higher coordination. However, at atomic-scale thicknesses, interlayer interactions have a reduced role, which can stabilize phases that are typically unstable in bulk such as single and few-layer A17 Sb (α-antimonene)[17]. Moreover, ultrathin A7 Bi(110) arranges in A17-like bilayers, even though an unambiguous identification of their stacking is still missing[18-23]. At a few bilayers, Bi(110) thin films rearrange to the (111) orientation[19-21]. Obviously, the physical properties are expected to emerge from the atomic structure and its evolution with thickness. Indeed, A17 P and As exhibit thickness-dependent direct bandgaps, high carrier mobilities, and anisotropic transport properties[3,24,25]. On the other hand, as their thickness approaches atomic scales, A7 As, Sb, and Bi undertake several electronic and topological transitions including topological insulator, topological semimetal, quantum spin Hall and semiconducting phases[26-28]. Understanding the stability limits of the A17 phase across the group VA and its transformation mechanism to the A7 phase is thus essential to control and harness the properties of this rich allotropic group of layered materials.

Herein, by studying in real-time the molecular beam epitaxial (MBE) vdW growth of antimony on epitaxial graphene using *in situ* low-energy electron microscopy (LEEM), we directly observed a hitherto unknown thickness-driven layered-to-layered phase transformation from the 2D A17 phase to the A7 phase. In the following section, β-antimonene refers to a single bilayer of A7 phase (Fig. 1a) or a stack of a few bilayers, whereas α-antimonene refers to a single bilayer of A17 phase (Fig. 1a). Multilayer α-antimonene can be AA or AB stacked and will be referred to as AA α-antimonene and AB α-antimonene. Bulk A17 Sb is an AB stack of α-antimonene bilayers. On the other hand, thick AA α-antimonene relaxes into the A7 structure with a (110) surface orientation and is referred to as A7 Sb(110). Our earlier studies demonstrate that β-antimonene can be grown on graphene[29,30]. Below, we show that under the same conditions, the deposition of Sb on graphene also leads to the additional growth of α-antimonene (AA and AB).

*in situ* LEEM of 2D-Sb on graphene (Fig. 1b) reveals β-antimonene and rectangular 2D islands. The *ex situ* μ-LEED pattern of a rectangular island (Fig. 1b (inset)) is similar to few-layer black phosphorus[31,32], with lattice parameters of 4.25 ± 0.05 Å and 4.57 ± 0.05 Å. The rectangular LEED pattern could realistically be associated with three structures: AB α-antimonene, AA α-antimonene, and A7 Sb(110). Bulk A7 Sb(110) has surface lattice parameters of 4.31 Å and 4.51 Å and the density functional theory (DFT) calculated in-plane lattice constants of few-layer AA α-antimonene and AB α-antimonene are $a_{AA}$=4.28 Å, $a_{AB}$=4.27 Å, $c_{AA}$=4.55 Å and $c_{AB}$=4.81 Å. We note that DFT calculations show that A7 Sb(110) rearranges into AA stacked α-antimonene at few bilayers thicknesses (Fig. S1). While this LEED pattern points towards AA stacked α-antimonene, large area *in situ* LEED tells another story (Fig. 1 (c)). With the ~100 μm$^2$ electron beam area, six diffraction rings are observed, indicating a random in-plane rotation of the islands due to their vdW interaction with graphene. The two smallest rings (4.7 ± 0.1 Å and 4.5 ± 0.1 Å) come from the (01) spots of AB and AA α-antimonene, respectively. The third ring (4.3 ± 0.1 Å) originates from the (10) spots of AB α-antimonene. The fourth ring (4.28 Å) is attributed to β-antimonene and the fifth and sixth rings are the (11) spots of AB and AA α-antimonene. As detailed below, the presence of AA α-antimonene/A7 Sb(110) is due to a phase transition when AB α-antimonene (A17 Sb) reaches a critical thickness at which it is no longer stable. Cross-sectional STEM of a ~2.1 nm thick AB α-antimonene 2D island on graphene/Ge viewed along the [101] direction is shown in Fig. 1d. The 0.7 nm gap between Ge and Sb indicates the presence of vdW interfaces between Ge/graphene and graphene/Sb. The bilayer structure of α-antimonene is clearly resolved and the average bilayer thickness and inter-bilayer distance are 2.7 ± 0.1 Å and 3.3 ± 0.1 Å, respectively. The corresponding DFT calculated values are 2.83 Å and 3.23 Å. The puckered α-antimonene structure and AB stacking of A17 Sb are also confirmed. The measured 3.2 ± 0.1 Å lateral periodicity agrees relatively well with the calculated 3.19 Å ($\bar{1}01$) d-spacing. Raman scattering spectroscopy recorded on different regions revealed vibrational spectra that are typical to β-antimonene with the two modes $E_g$ ~124 cm$^{-1}$ and $A_{1g}$ ~150 cm$^{-1}$ (Fig. 1f). Additionally, spectra with three Raman modes at 133.7 cm$^{-1}$, 146.7 cm$^{-1}$ and 160.6 cm$^{-1}$ were also observed (Fig. 1f), thereby confirming the presence of AB α-antimonene. In analogy to the well-studied A17 black phosphorus[33], these three peaks are attributed to the $A_g^1$, $B_{2g}$ and $A_g^2$ modes, respectively.

Next, we analyze the growth behavior of α-antimonene using *in situ* LEEM (Fig. 2 (a-f)). AB α-antimonene islands usually nucleate at defects on the graphene surface (Fig. S2) and grow laterally, with their long axis aligned with the [100] direction (Fig. 2 (a, b) and Fig. S2). During this growth stage, the islands reach lengths of ~1 µm and widths of ~0.2-0.5 µm. A sudden and homogeneous decrease in LEEM intensity then occurs over the whole island faster than LEEM data acquisition speed (1 frame/s) (Fig. 2c and supplementary video 1). Then, the 2D island grows laterally only along its long axis (Fig. 2 (c-f)) and nanowires form on each of its long edges. Simultaneously, linear nanostructures aligned with the long axis appear on the 2D island's surface, leading to additional nanowire nucleation in the center (Fig. 2f). A model of the evolution α-antimonene during growth is proposed in Fig. 2g. The islands first nucleate in the AB α-antimonene (A17) phase, which is stabilized by the nanoscale thickness (LEEM in Fig. 2 (a, b)). In fact, DFT indicates that few layer AB α-antimonene is more stable than few-layer A7 Sb(110)/AA α-antimonene (Fig. 2h). However, a stability crossover between the two phases occurs at 7 bilayers (~4 nm), allowing the phase transformation to AA α-antimonene. In fact, the rate of the homogeneous decrease in LEEM intensity (Fig. 2c) rules out the growth of an overlayer on the island, which would occur in a timescale of a few seconds at these dimensions. The subsequent evolution is explained by the crystal structures and stabilities of A7 Sb(110) and AA α-antimonene (Fig. S3). In fact, A7 Sb(110) is related to AA α-antimonene by a shear deformation $\gamma_{xy}$ in the a-b plane of the orthorhombic unit cell (Fig. S3) into a monoclinic supercell (Fig. 2g) with a thickness dependent $\beta=86.4-90°$ at thicknesses above 12 bilayers (~7.4 nm). While DFT indicates that thin A7 Sb(110) films relax into AA α-antimonene (Fig. S3), thicker films have stronger inter-bilayer bonds, and weaker intra-bilayer bonding (Fig. 2g and Fig. S3), leading to a gradual transition of the bilayer structure from the (110) planes to the (111) planes as the thickness increases. The (111) bilayers stabilization in thicker films is at the origin of the facetted nanowires and nanostructures observed in LEEM after the phase transition (Fig. 2 (d-f)).

High-resolution STEM studies (Fig. 3(a-d)) unravel more details on the phase transition and nanowire formation mechanisms. STEM along the [1$\bar{1}$0] and [$\bar{1}$1$\bar{1}$] directions confirms that the 2D islands transition to A7 Sb(110) (Fig. 3a and Fig. S4). Interestingly, rotated grain boundaries are seen at the interface between the nanowires and the island (Fig. 3b). The nanostructures observed in LEEM are 1D nano-ridges aligned along the island's long axis ([1$\bar{1}$0] direction). μ-LEED also confirms the presence of 1D nano-ridges (supporting video 2). Twin boundaries ((112) planes) are found at each valley and hill of the nano-ridges structure (Fig. 3 (a, c, d)), ultimately allowing facets to tend towards the low-energy (111) planes. AFM indicates that the twinning leads to a 1D array of nano-ridges with a ~35 nm periodicity on the islands surface (Fig. 3e). The twin domains formation strongly support a AB to AA α-antimonene shuffle transition mechanism followed by relaxation to A7 Sb(110). In fact, both twin domains can be obtained from the intermediate AA α-antimonene structure, by a positive or negative $\gamma_{xy}$ shear deformation (Fig. S5). This suggests that the islands are in the AA α-antimonene phase just after the transition, when no nanostructure is observed in LEEM (Fig. 2c). Relaxation to bulk-like A7 Sb then occurs in thicker islands by the formation of twin domains and (111) facets, driven by the minimization of surface energy. In fact, AFM analysis indicates that thin islands (< 4.5 nm) have flat surfaces, whereas most of the thicker ones (>4.5 nm) display a nano-faceted surface (Fig. S6), in good agreement with the A17 stability limit determined by DFT and suggesting that the twin domains formation occurs shortly after the phase transition.

The twin domain formation by shear deformation is made possible by the hybrid bonding structure in AA α-antimonene and A7 Sb(110) thin films (Fig. 4), where both α-antimonene and bulk A7 bonding features are present. In addition to the 3 in-plane covalent bonds in the (111) bilayers (Fig. 4a), the atoms in thin Sb(110) islands have an additional fourth interlayer bond (Fig. 4b), which belongs to the α-antimonene bilayer structure. This makes the positive and negative $\gamma_{xy}$ shear deformations equivalent and without significant energy barriers. The analysis of the XPEEM spectra of thin A7 Sb(110) and Sb(111), substantiated by DFT calculations, supports the existence of this intermediate phase (Fig. 4 (c-e)), as discussed below. The (Sb $4d_{5/2}$, Sb $4d_{3/2}$) binding energies are (32.19 eV, 33.42 eV) for thin Sb(110) and (32.45 eV, 33.58 eV) for thin Sb(111) (Fig. 4e). Since the Sb 4d peaks can be fitted with a single component (FWHM=0.6 eV), the 160 meV XPS shift is attributed to electronic structure differences between the two pristine phases. The trends in the calculated density of states (DOS) allow to better interpret the observed shifts (Fig.

4f). The DOS shows 47 meV and 336 meV Sb 4d binding energy red shifts in 8 bilayers Sb(110) and bulk A17 Sb, with respect to bulk A7 Sb. While this qualitative agreement supports the presence of an intermediate phase, the Sb(110) valence electrons' DOS closely resembles the bulk A7 Sb DOS. However, the structure of thin Sb(110) films has α-antimonene features (Fig. S7). In fact, bulk A7 Sb atoms have three nearest neighbors (NNs) (2.92Å) and three second NNs (3.28 Å). Bulk A17 Sb atoms have one NN (2.87Å) and a two second NNs (2.90Å). For thin Sb(110) films (12 bilayers), the surface layers are α-antimonene-like and the middle layers have A7 and A17 features, with two NNs (2.90Å) and almost equivalent third and fourth neighbors (3.08Å and 3.10Å ) (Fig. S7). This intermediate bonding structure is due the competition between Sb(110) surface bilayer reconstruction and bulk (111) bilayer formation.

Significant interlayer bonding in AA α-antimonene begins at 4 bilayers thicknesses, especially between the middle bilayers (Fig. S8). These interlayer bonds probably have an important role in the nucleation of AA α-antimonene. In fact, it was suggested that the high pressure-induced A17 to A7 phase transition in bulk phosphorus nucleates by the formation of interlayer bonding chains during shuffling, followed by the growth of interlayer bonding regions[15]. This suggests that AA α-antimonene nucleates on multiples layers at once, especially since DFT shows that stacking faults are not energetically favorable. However, classical nucleation theory indicates that homogeneous nucleation of AA α-antimonene is unlikely, as it predicts critical nuclei sizes of $r^* = 18\ Å$ and nuclei surface energies of $\gamma = 0.64\ meV/Å^2$ to explain the observed nucleation rate (details in supporting information). This very low $\gamma$ suggests heterogeneous nucleation of AA α-antimonene, perhaps at the islands' edges. We also note that the lattice mismatch-induced ~5% uniaxial strain along the c axis in the nucleus can be more easily accommodated by the A17 α-antimonene lattice if nucleation occurs on the edges of the quasi-free-standing island, rather than at the center.

In summary, by addressing the unique thermodynamic behavior of layered materials we demonstrated the co-existence of metastable A17 phase and stable A7 phase of Sb during the vdW growth on weakly interacting surfaces. The ultrathin A17 Sb phase (AB α-antimonene) is identified by LEEM/LEED, STEM and Raman spectroscopy. *in situ* LEEM growth observations show that the A17 phase is stable at thicknesses below ~4 nm. Growth beyond this thickness yields a spontaneous transition from the A17 phase to the A7 bulk-like phase. We proposed a diffusionless shuffle phase transition mechanism from AB α-antimonene to an intermediate AA α-antimonene with both A17 and A7 bonding and electronic features. The intermediate phase eventually evolves towards bulk-like A7 by the formation of twin domains and periodic 1D arrays of nano-ridges with (111) facets. This phase transition mechanism is supported by STEM and *in situ* LEEM observations. These results highlight the complex interplay between the atomic structure and stability of layered materials and lay the groundwork for phase engineering in group VA elements.

**Acknowledgements**

O.M. acknowledges support from NSERC Canada (Discovery grants, Strategic Partnership grants, Collaborative Research and Development Grants), PRIMA Québec, Canada Research Chairs, Canada Foundation for Innovation, Calcul Québec. The electron microscopy work was carried out at the Canadian Centre for Electron Microscopy, a national facility supported by the Canada Foundation for Innovation under the MSI program, NSERC and McMaster University. R.M.J. and M.S.A. acknowledge support from the U.S. Department of Energy, Office of Science, Basic Energy Sciences, under Award # DE-SC0016007 for graphene synthesis. S.U., J.A.M. and D. B. acknowledge financial support from Villum Fonden (grant no. 15375), from the Danish Council for Independent Research, Natural Sciences under the Sapere Aude program (grant no. DFF-6108-00409), and from Aarhus University Research Foundation. This research used resources of the Advanced Light Source, which is a DOE Office of Science User Facility under contract no. DE-AC02-05CH11231. Computations were made on the supercomputers Beluga, Cedar and Graham, managed by Calcul Québec, WestGrid, Compute Ontario and Compute Canada.

**Author contributions**

M.F.-D. and O.M. conceived the experiments. M.F.-D. performed the MBE growth experiments, Raman AFM, LEEM/LEED, XPEEM analyses, data processing, and DFT calculations. R.M.J. and M.S.A. provided epitaxial graphene substrates. H.Z., M.J.L. and G.A.B. performed STEM studies. T.O.M., A.L. and F.G. led the LEEM and XPEEM work at Elettra. D.B., C.J., J.A.M., S.U., A.B., and E.R. contributed to the MBE growth and LEEM studies at the Advanced Light Source. M.F.-D. and O.M. wrote the manuscript and all authors commented on it.

**Methods**

*MBE growth.* Antimonene was grown under ultra-high vacuum (UHV) (P<1E-9 mbar) by solid source molecular beam epitaxy (MBE) on epitaxial graphene on Ge(110) by deposition of $Sb_4$ species using a Knudsen cell. Sb crystals with 99.9999% purity were used as $Sb_4$ source. The substrate temperature was kept at constant temperatures between 190-250 °C during growth and the deposition rates between 1.5-25 nm/min (calibrated at room temperature on Ge substrates; effective growth rates are lower due to $Sb_4$ desorption). Graphene was grown on Ge substrates by chemical vapor deposition following the process described in Refs. [34,35]. Prior to growth, the graphene substrates were annealed at 600°C for ~10 min under UHV.

*in situ LEEM characterization.* The MBE growth was carried out under low-energy electron microscopy (LEEM) observations using a FE-LEEM P90 from SPECS GmbH. The LEEM data acquisition rate was 1 frame/second. After growth, low-energy electron diffraction (LEED) was measured *in situ* using a ~100 µm² electron beam to determine the crystal structure of the grown antimonene phases. The (10) LEED spots of β-antimonene were used for reciprocal space calibration using the previously determine 4.28 Å lattice constant[29]. Additional MBE growths and post-growth LEEM characterization of the samples were carried out at the MAESTRO beamline of the Advanced Light Source in Berkeley National Laboratory.

*Scanning transmission electron microscopy characterization.* After growth, samples were taken out in air for *ex situ* characterization. Transmission electron microscopy samples were prepared from selected flakes by a cross-section lift out and thinning to electron transparency using a focused ion beam (Zeiss NVision 40 FIB). Imaging by scanning transmission electron microscopy (STEM) was carried out using a FEI Titan 80-300 Cubed TEM, equipped with a XFEG source and CEOS-designed hexapole-based aberration correctors for both the probe-forming lens and the image lens. STEM measurements were performed at 200 kV with a semi-convergence angle of 19 mrad. The signal was acquired on a high angle annular dark field (HAADF) detector with an inner acceptance semi-angle of 64 mrad. Images were averaged from frames which were rigidly and non-rigidly aligned using Smartalign software in order to reduce noise of the scanning unit.

*X-ray photoemission electron microscopy characterization.* X-ray photoemission electron microscopy (XPEEM) and µ-LEED were carried out at the Nanospectroscopy beamline at the Elettra Synchrotron laboratory. XPEEM was carried out with an incident X-ray beam (250 eV) at 16° incidence angle[36]. The photoelectron energy resolution is better than 300 meV, and the XPEEM spatial resolution is better than 30 nm[37].

*Raman scattering spectroscopy characterization.* Raman scattering spectra were acquired using a Renishaw inVia system with 633 nm laser excitation on samples exposed to atmospheric conditions for less than one week.

*Atomic force microscopy characterization.* Atomic force microscopy images were acquired in tapping mode using a Veeco Dimension 3100 microscope equipped with a silicon probe on samples exposed to atmospheric conditions for less than two weeks. WsXM was used for AFM data analysis[38].

*Density functional theory calculations.* Density functional theory calculations (DFT) were carried out with Quantum Espresso[39] using the projector-augmented wave method[40] within the generalized-gradient approximation (GGA). The PBE[41] functional was used and van der Waals interactions were accounted for with Grimme dispersion correction (DFT-D2)[42]. Slab geometries with at least 15 Å vacuum separating periodic images were used. (1×$N$×1) orthorhombic supercells were used for AA and AB stacked α-antimonene, with $N$ the number of bilayers going from 1 to 14. (1×$N$×1) monoclinic supercells were used for A7 Sb(110) with $N$ going from 1 to 20. For cohesive energy calculations, Sb 4d electrons were treated with the frozen core approximation and a 517 eV cut-off was used for plane-wave expansion along with a (9×1×9) Monkhorst-Pack k-point mesh[43]. For density of states calculations, Sb 4d electrons were treated as valence electrons and a 748 eV cut-off was used for plane-wave expansion along with a (9×1×9) Monkhorst-Pack k-point mesh. The convergence criteria for electronic and structural relaxation were set to 1μeV and 0.01eV/ Å, respectively.

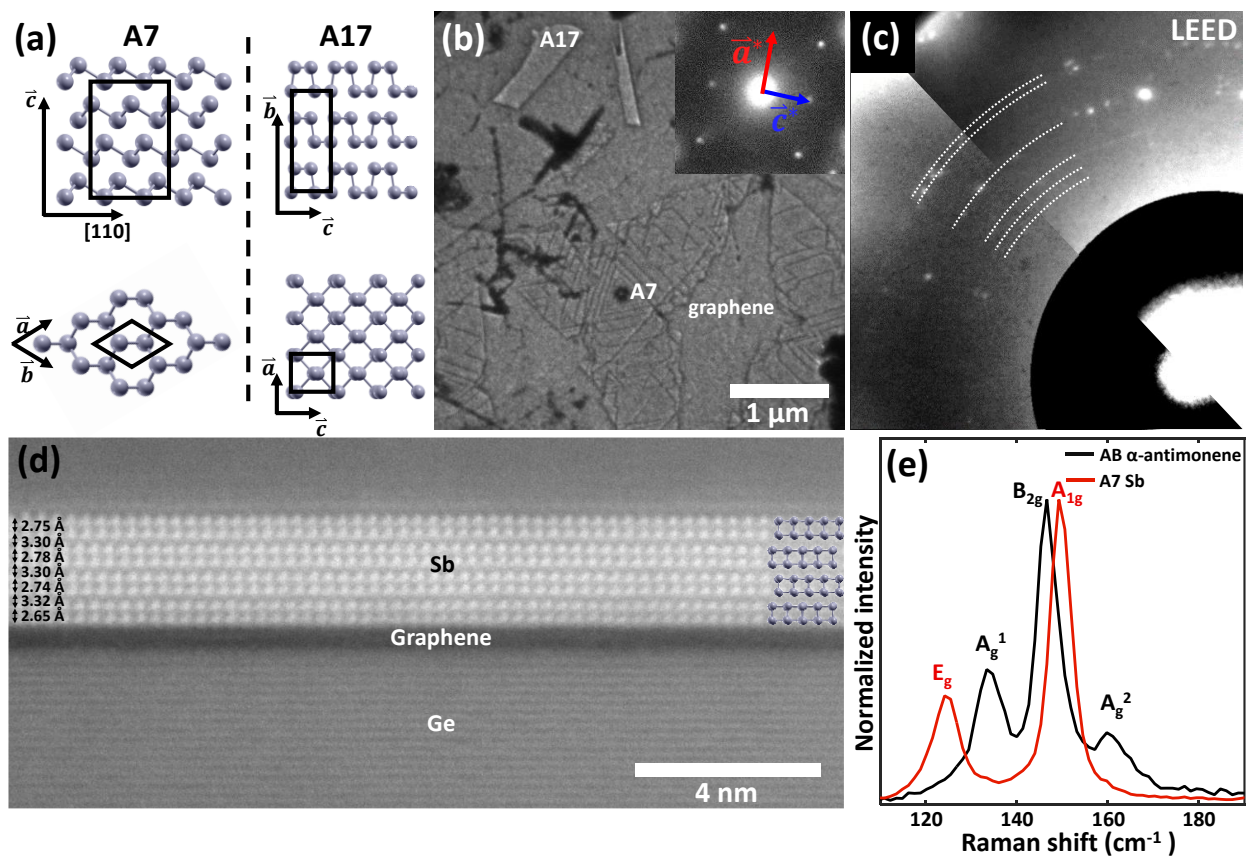

**Figure 1 Identification of A17 2D-Sb** (a) Atomic structure of A7 (left) and A17 Sb (right). On top are the side views of the vdW layered structures viewed along the [1$\bar{1}$0] and [100] directions, respectively and on the bottom are the top views of the atomic structures of individual bilayers. (b) Bright-field LEEM (2.2 eV) of 2D-Sb grown on epitaxial graphene on Ge(110). A μ-LEED pattern of a single rectangular island (AA α-2D-Sb) is shown in the inset. (c) Large area LEED pattern of 2D-Sb on graphene. To allow to view the six diffraction rings, 1/8 of the Ewald sphere at 44 eV is shown in the bottom left and 1/8 at 29 eV is shown in the top right. (d) Cross-sectional STEM of 4 bilayers thick AB α-antimonene island on graphene viewed along the [101] direction. (e) Raman spectra of AB α-antimonene and β-antimonene on graphene.

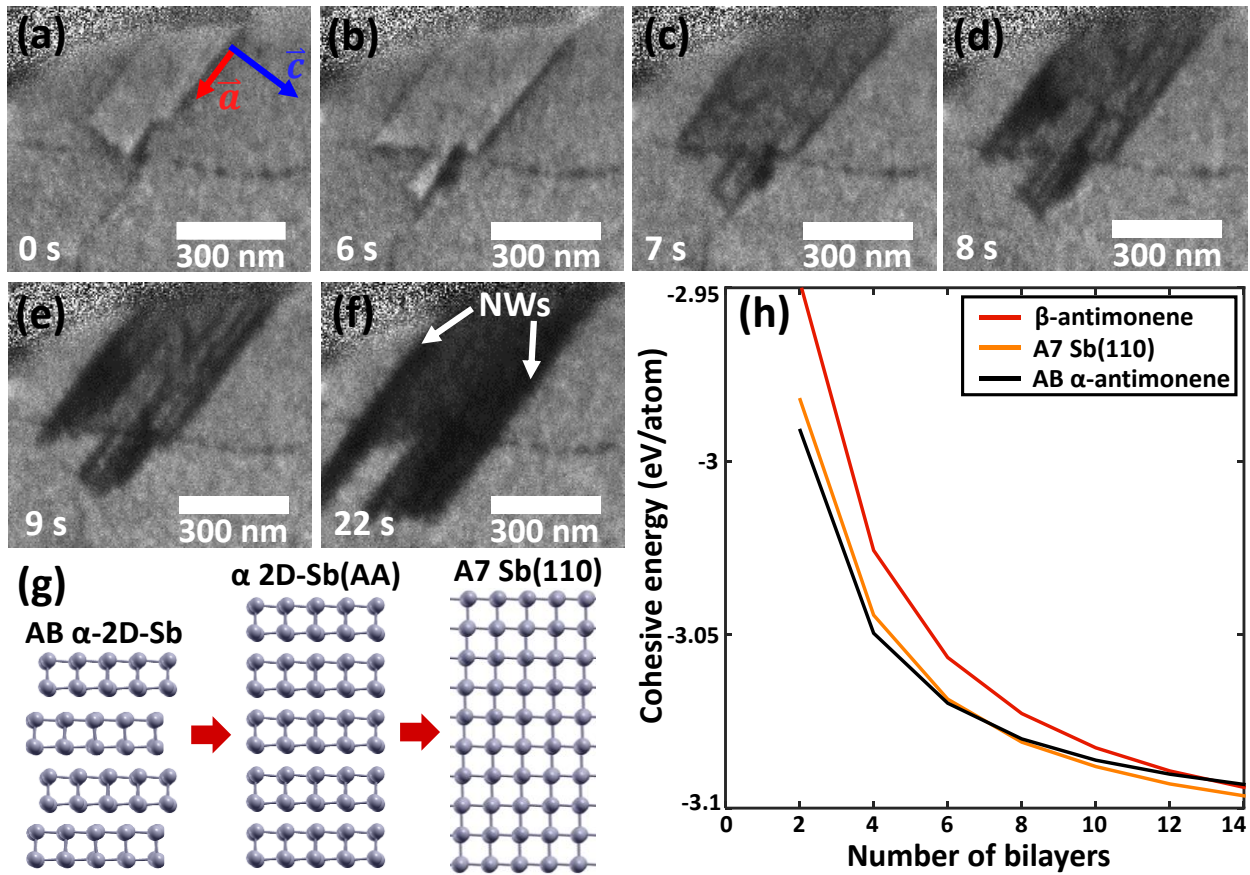

**Figure 2 Growth and phase transition behavior of A17 2D-Sb** (a-f) *in situ* bright-field LEEM (2 eV) of AB α-antimonene growth and its transition to A7 Sb(110) when it reaches its critical thickness. Deposition rate is 16 nm/min and substrate temperature is 230 °C (g) Phenomenological model of A17 → A7 Sb phase transition. (h) DFT calculated cohesive energies for the 2D-Sb allotropes.

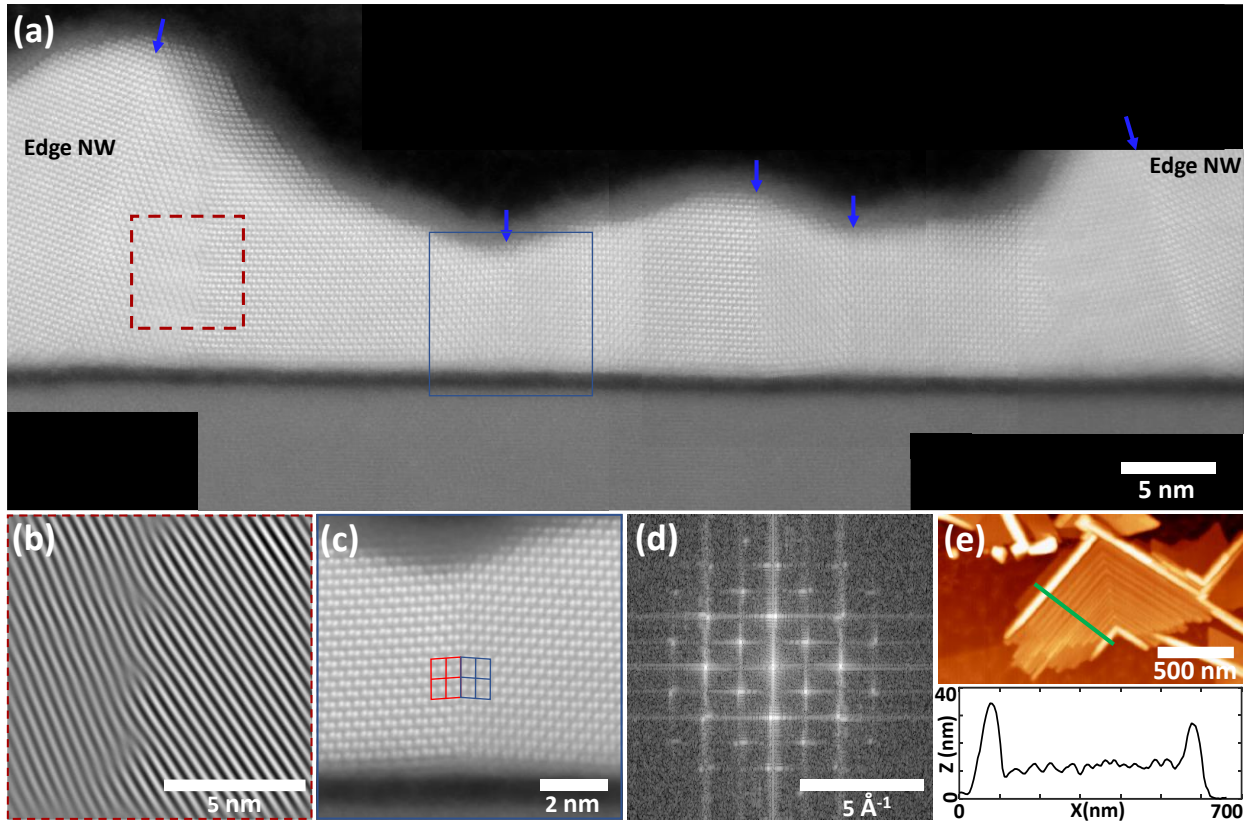

**Figure 3 Characterization of thin Sb(110) islands** (a) STEM images of A7 Sb(110) viewed along the $[1\bar{1}0]$ direction. Blue arrows indicate the twin boundaries. (b) FFT filtered image ((111) planes) from the region in the red dashed square in Fig. 3a. (c) High-magnification STEM from the region in the blue square in Fig. 3a. Monoclinic supercells for each twin domain at the boundary are shown in red and blue. (d) FFT from the Sb region in Fig. 3c. (e) AFM from a A7 Sb(110) island showing a nano-ridges structure and nanowires at the edges. Here, the two branches of the L-shaped island have an in-plane crystallographic orientation rotation of 87° with respect to each other.

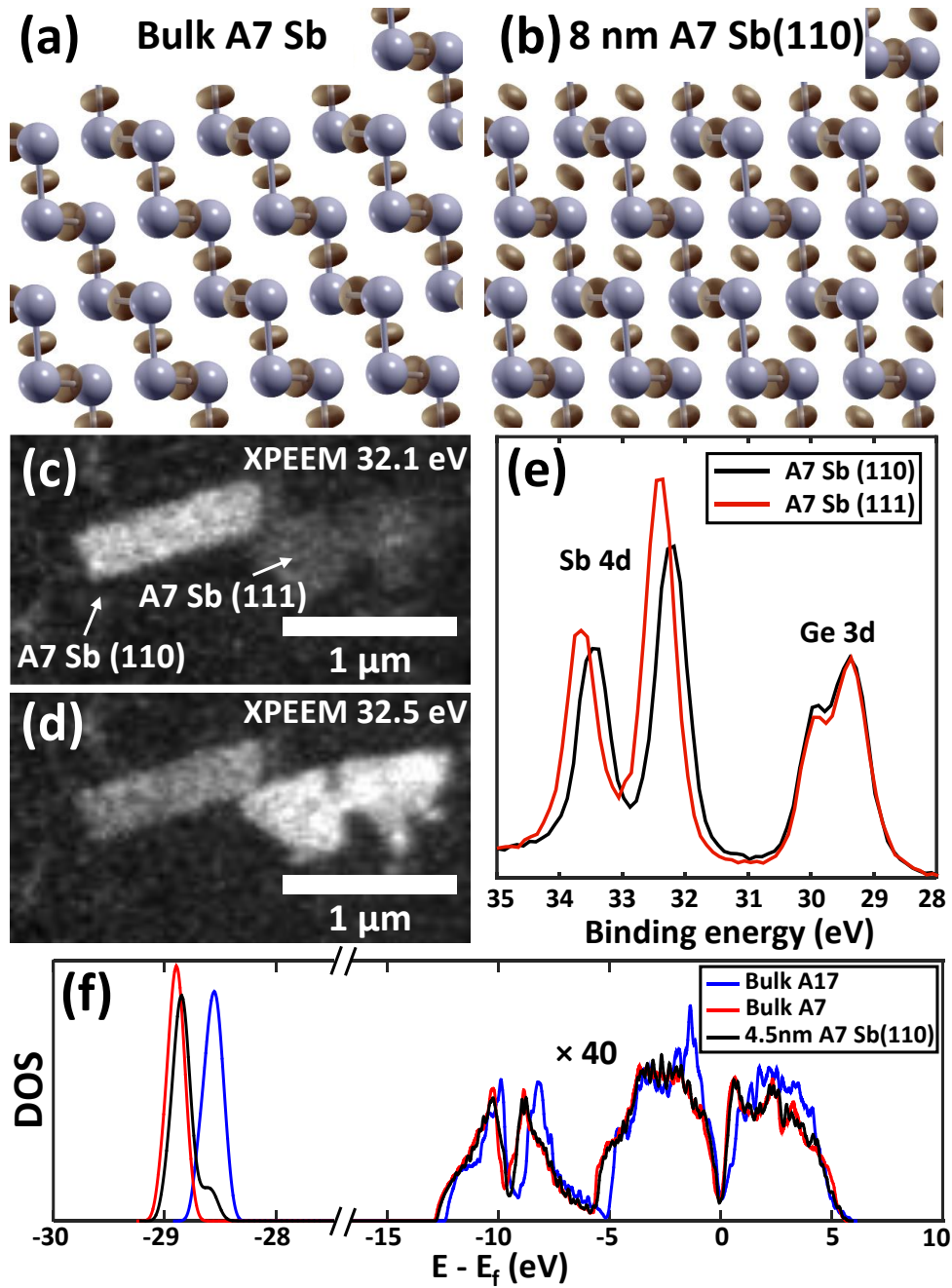

**Figure 4 Electronic and bonding structure of 2D-Sb phases** (a, b) DFT calculated charge density minus superposition of atomic charge densities iso-surface at 0.0035 e$^-$/a$_0^3$ for: (a) bulk A7 Sb and (b) 14 bilayers thick A7 Sb(110) film. (c, d) XPEEM images using 250 eV photon energy at: (c) 32.1 eV binding energy and (d) 32.5 eV binding energy (supporting video 3). The rectangular island is thin A7 Sb(110) and the triangular island is thin A7 Sb(111). (e) Corresponding XPS spectra extracted from XPEEM shown in Fig. 4 (c, d). (f) DFT calculated density of states (DOS) for bulk A17 Sb, bulk A7 Sb and 8 bilayers thick A7 Sb(110). The x-axis scale is different between [-30, -28] eV and [-15, 10] eV to allow visualization of d and s-p states.

# Supporting Information to: Pnictogens Allotropy and Phase Transformation During van der Waals Growth

Matthieu Fortin-Deschênes[1], Hannes Zschiesche[2], Tevfik Onur Menteş[3], Andrea Locatelli[3], Robert Jacobberger[4], Francesca Genuzio[3], Maureen J. Lagos[2], Deepnarayan Biswas[5], Chris Jozwiak[6], Jill A. Miwa[5], Søren Ulstrup[5], Aaron Bostwick[6], Eli Rotenberg[6], Michael Scott Arnold[4], Gianluigi A. Botton[2], Oussama Moutanabbir[1]

[1] Department of Engineering Physics, École Polytechnique de Montréal, C. P. 6079, Succursale Centre-Ville, Montréal, Québec, H3C 3A7, Canada

[2] Department of Materials Science and Engineering, McMaster University, Hamilton, Ontario, L8S 4L8, Canada

[3] Elettra Sincrotrone Trieste S.C.p.A, S.S. 14 – km 163, 5 in AREA Science Park, 34149 Basovizza, Trieste, Italy

[4] Department of Materials Science and Engineering, University of Wisconsin-Madison, Madison, Wisconsin 53706, USA

[5] Department of Physics and Astronomy, Interdisciplinary Nanoscience Center, Aarhus University, 8000 Aarhus C, Denmark

[6] Advanced Light Source, Lawrence Berkeley National Laboratory, Berkeley, CA 94720, USA

The relaxed (DFT) atomic structures of bulk A7 Sb(110) and 4 bilayers thick Sb(110) are shown in Figure S1. Bulk Sb forms bilayers, which are aligned with (111) planes. On the other hand, the bilayers in thin Sb(110) films are oriented in the (110) planes. Moreover, we notice that the atomic structure of few bilayers thick Sb(110) films and few-layer AA α-antimonene are identical.

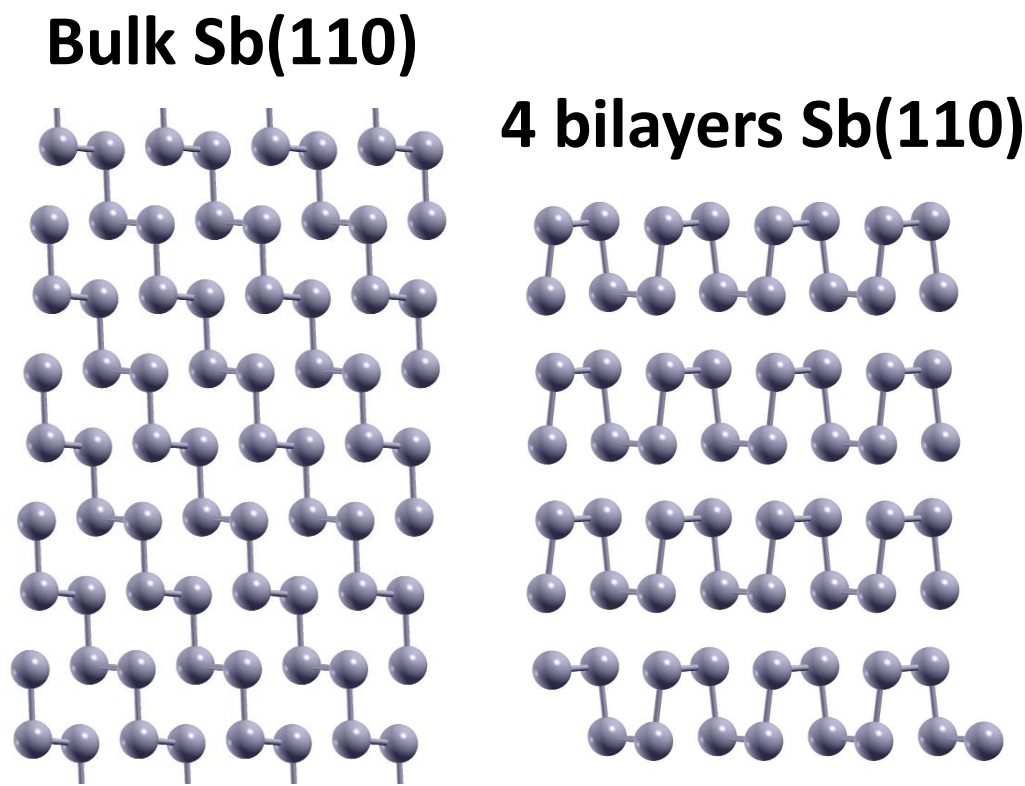

**Figure S5** Structure of bulk A7 Sb(110) (left) and DFT relaxed 4 bilayers thick Sb(110) film (left)

LEEM snapshots of the nucleation and growth of α-antimonene are shown in Figure S2. The bright rectangular islands are AB α-antimonene, whereas the dark rectangular islands are AA α-antimonene/A7 Sb(110). Nucleation of AB α-antimonene (white circles) can be seen at defects in the graphene surface. Only defect assisted nucleation is observed, which is due to the small adsorption energy of $Sb_4$ on graphene and the associated low surface concentration of precursor species.

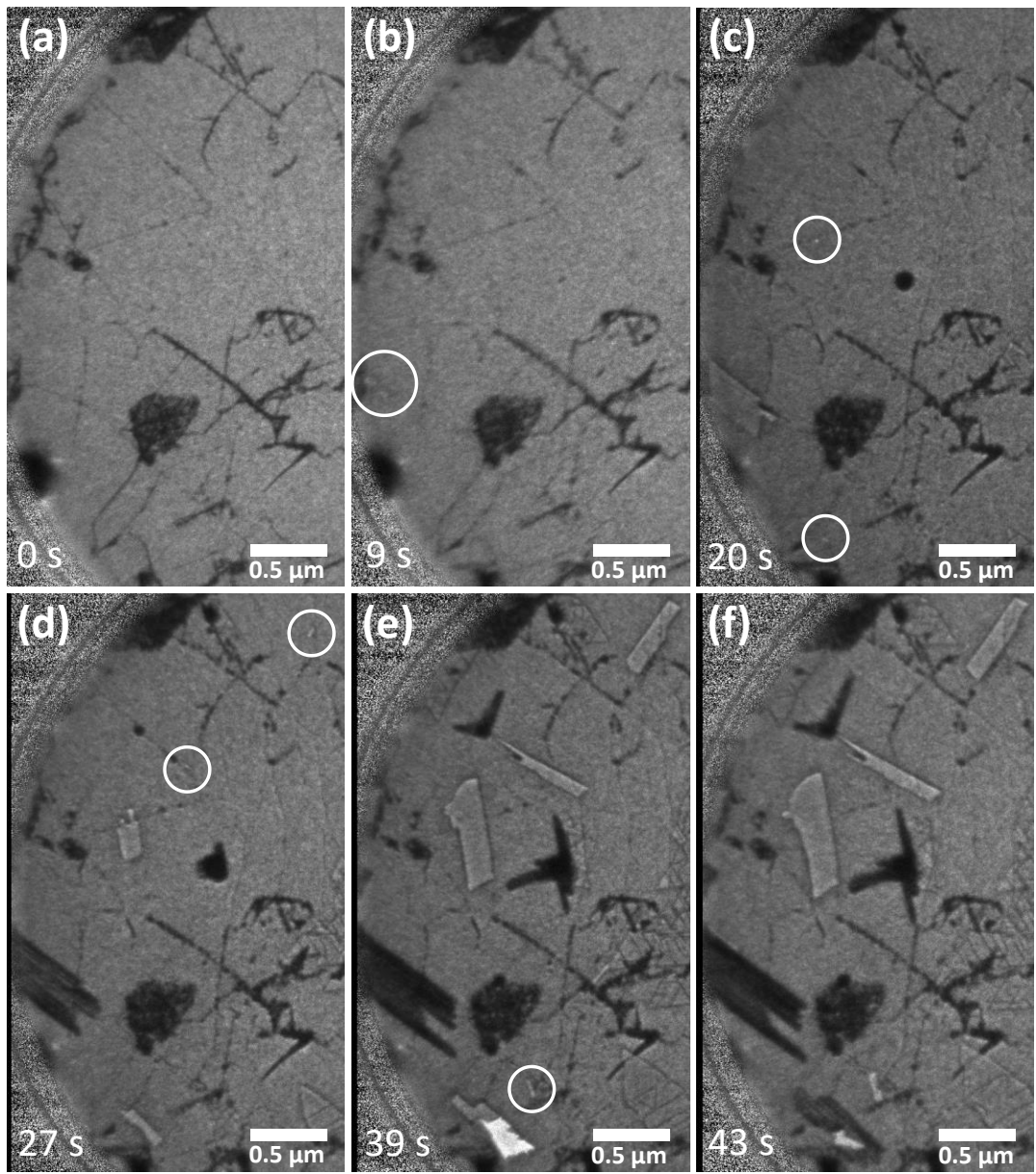

**Figure S6** Bright-field LEEM (2 eV) observations of A17 α-antimonene growth on epitaxial graphene on Ge(110). Deposition rate (calibrated at room temperature) is 16nm/min and substrate temperature is 230°C. A17 α-antimonene nuclei are circled in white and nucleation is observed to occur mostly at graphene defects.

The DFT relaxed structures of bulk A7 Sb(110) and thin A7 Sb(110) films is shown in Figure S3. The gradual transition of the bilayer orientation (110) to (111) can be seen as the thickness increases. In 14 bilayers film, clear (111) bilayers are observed in the center of the film.

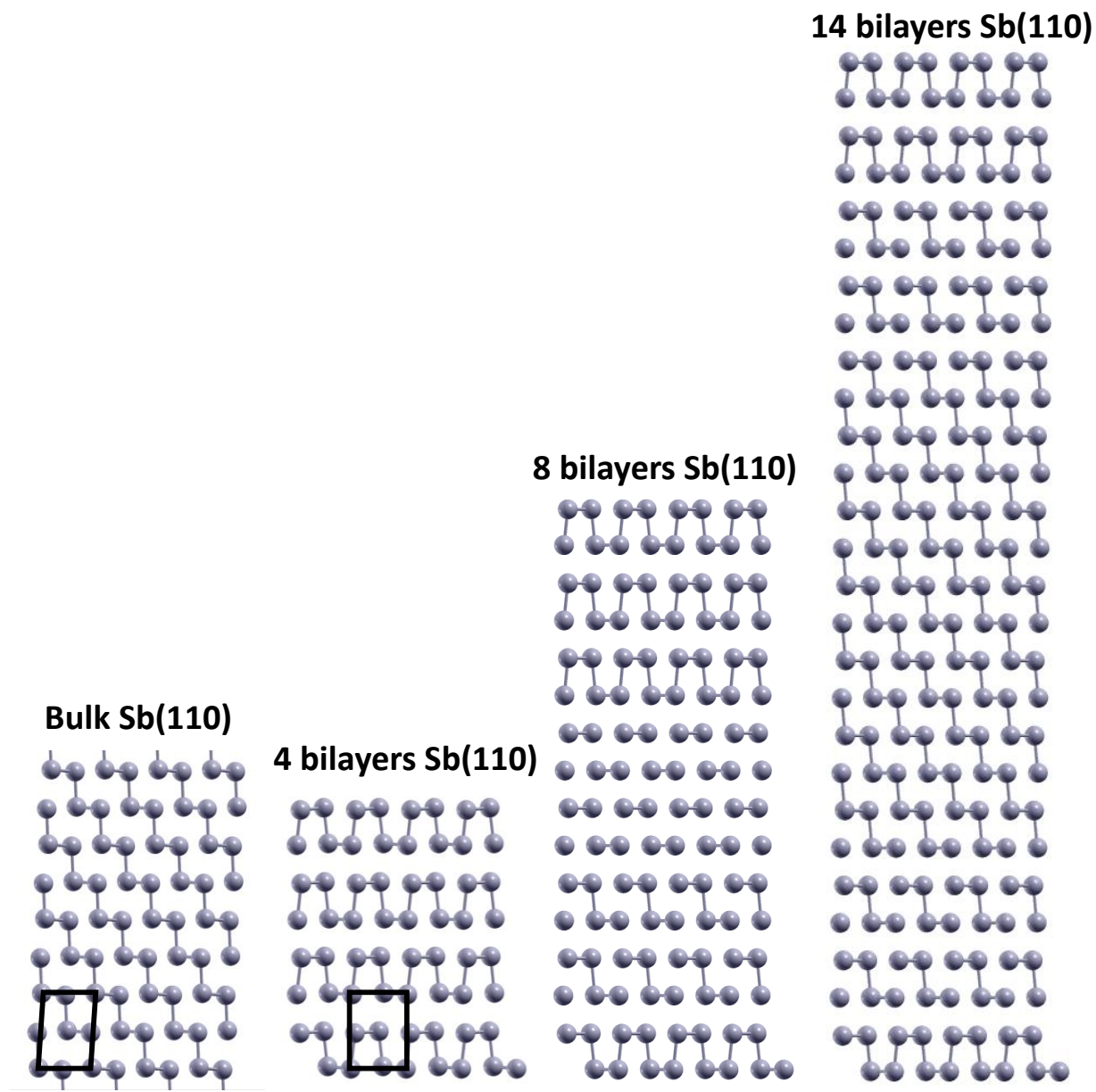

**Figure S7** DFT relaxed structures of bulk A7 Sb and Sb(110) thin films. Thin films rearrange in an AA α-antimonene bilayer structure. Thicker films begin to form stronger inter-bilayer bonds

and the bilayer structure gradually changes from the (110) planes to the (111) planes as thickness increases.

The STEM image in Figure S4 ($[\bar{1}1\bar{1}]$ view) confirms that the nanostructured islands are indeed A7 Sb(110).

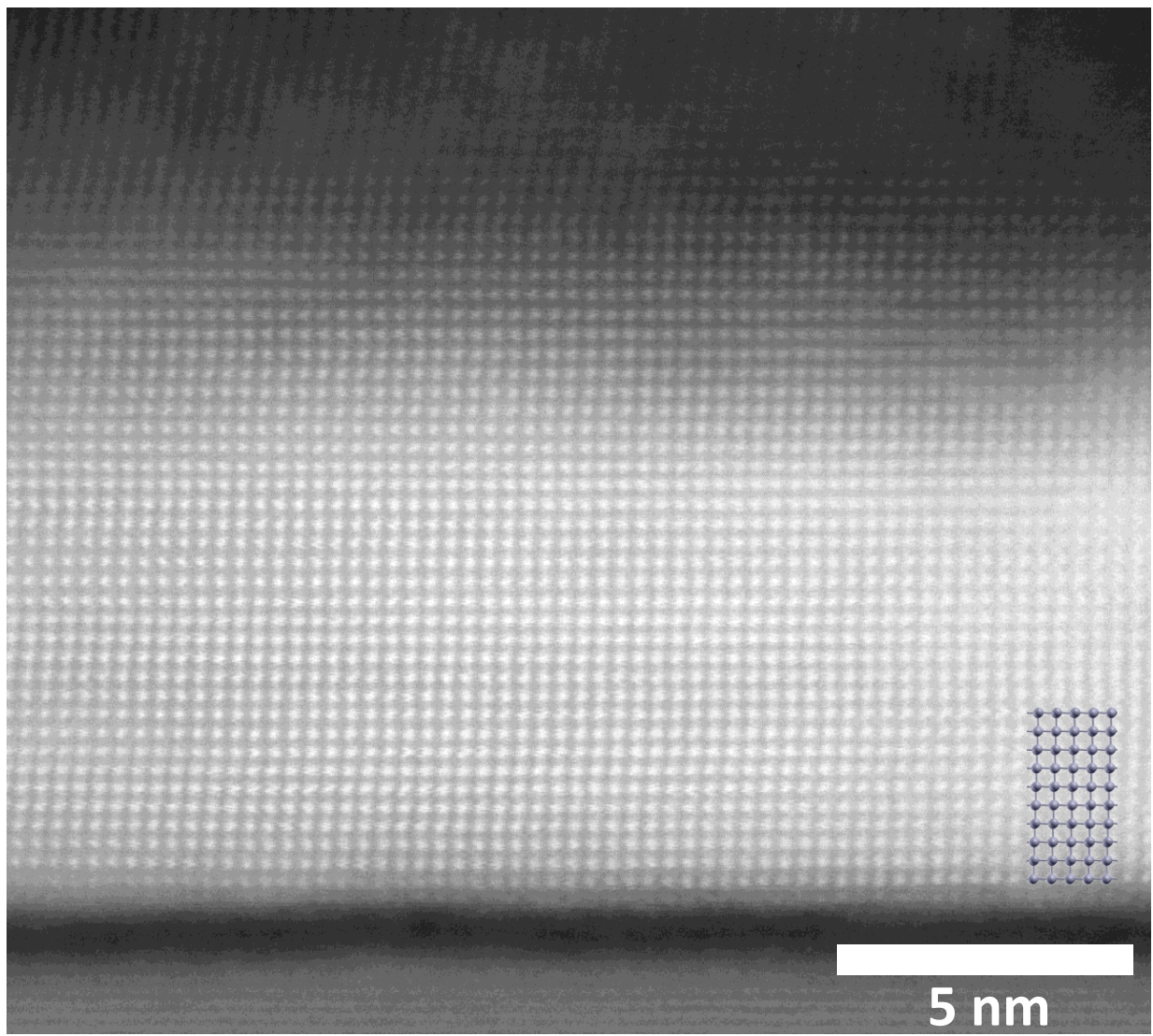

**Figure S8** STEM image of nanofaceted Sb(110) island viewed along the $[\bar{1}1\bar{1}]$ direction.

The presence of facetted nanostructures on the islands after the phase transition is confirmed by STEM and µ-LEED (supporting video 2). In fact, LEED I-V on single islands reveal the presence of multiple LEED spots which move along the $[001]$ and $[00\bar{1}]$ reciprocal direction as the kinetic energy of the incident electrons changes. This indicates that the surface of the island is not perpendicular to the incident electron beam and is tilted with its normal in the [01*X*] direction. The value of *X* cannot be determined since the facets are not fully formed and therefore not along a specific crystal plane. This interpretation is supported by the asymmetry of the LEED patterns.

The phase transition mechanisms from the intermediate AA α-antimonene phase to bulk-like A7 Sb(110) is illustrated in Figure S5. Positive or negative shear of the orthorhombic unit cell of AA α-antimonene leads to the formation of either of the two Sb(110) twin domains observed in STEM with a monoclinic supercell.

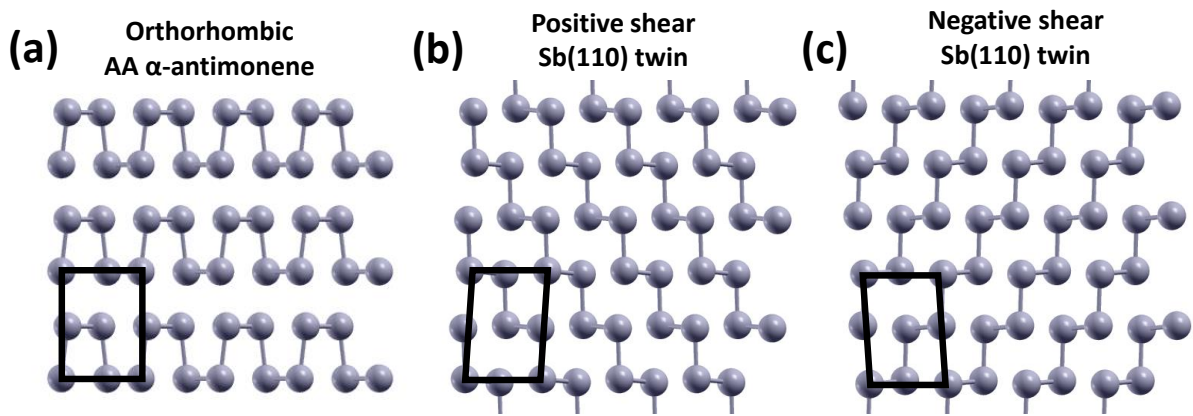

**Figure S9** (a) Structure of AA α-antimonene thin film with an orthorhombic unit cell. (b, c) DFT relaxed structures after applying a positive (a) or negative (b) shear on bulk orthorhombic AA Sb. The possibility to relax in either direction is at the origin of the twin domains seen in thin Sb(110) islands.

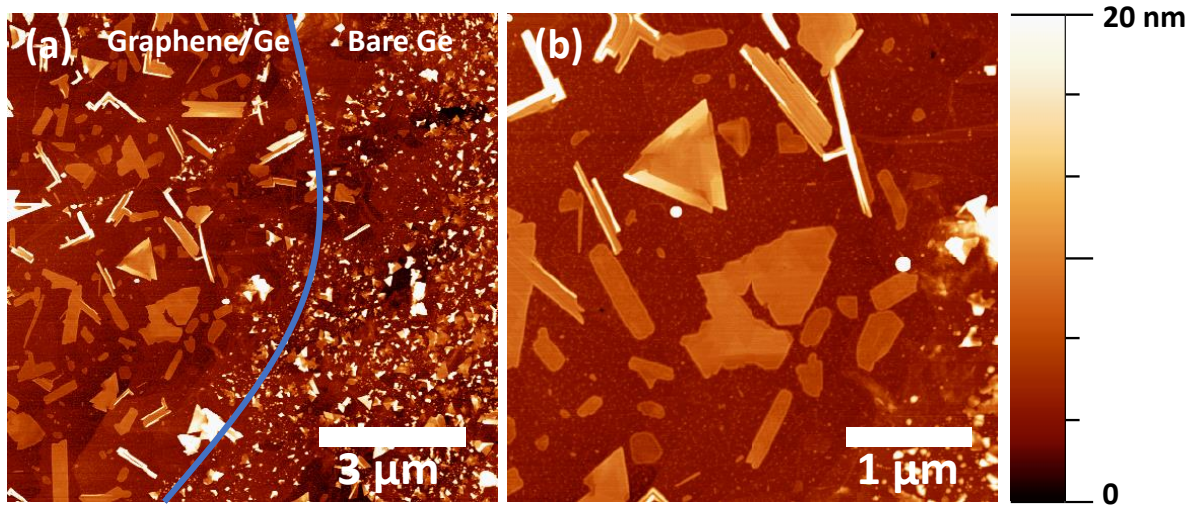

**Figure S10** AFM images of Sb deposited on epitaxial graphene on Ge(110). We note a large hole in the graphene layer (exposed Ge) at the right side of panel (a).

Using the DFT relaxed atomic structures of Sb(110) thin films, we computed the nearest-neighbor distances as a function of the position of the atom in the film (Figure S7). At a thickness of 12 bilayers, the hybrid A17/A7 bonding structure is observed throughout the film. The surface bilayers have a mostly α-antimonene structure, whereas the atoms in the bulk of the film form two short bonds in the (110) plane and two hybrid bonds, which allow the relaxation to either of the two twin domains observed in A7 Sb(110) islands. At 18 bilayers, the surface bilayer still has the α-antimonene structure. On the other hand, as we go closer to the center of the film, the structure tends towards the bulk-like A7 phase.

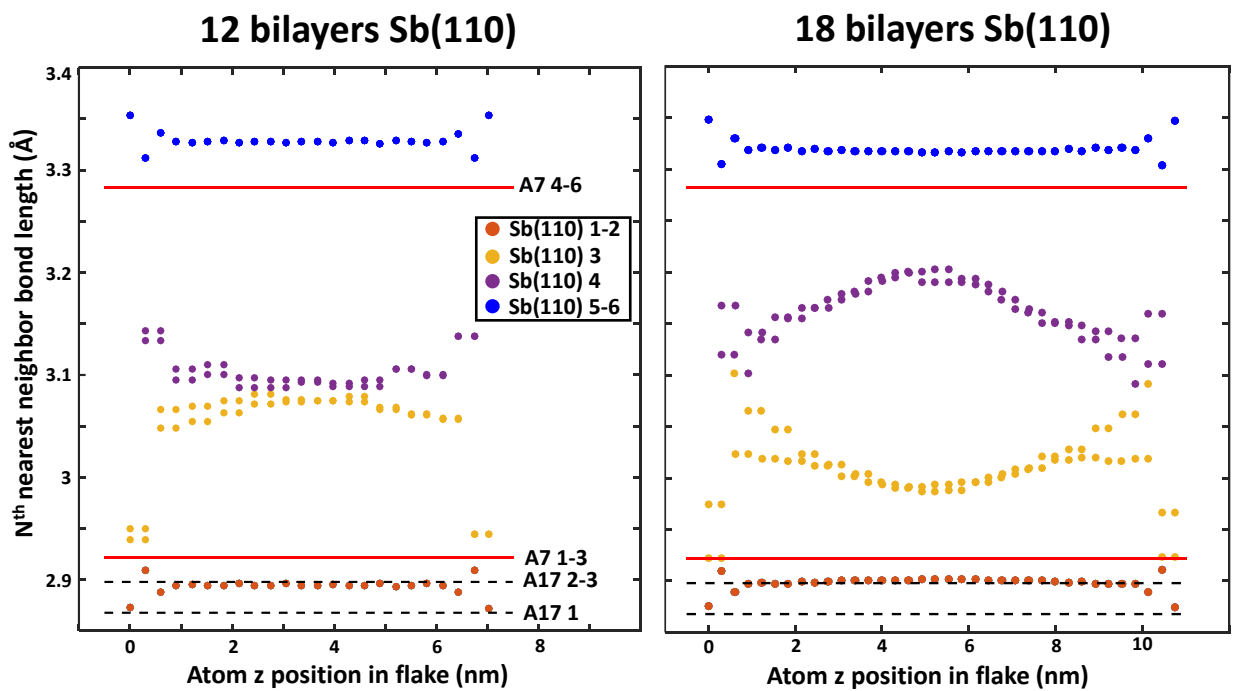

**Figure S11** Nearest-neighbors (NN) bond lengths in DFT relaxed Sb(110) films (12 and 18 bilayers thick). The two dashed black lines are the NN bond lengths for first NN and second NN in bulk A17 Sb. The two red lines are the NN bond lengths for first NN and second NN in bulk A7 Sb.

The hybrid bonding structure of thin AA α-antimonene is shown in Figure S8. The charge density minus superposition of atomic charge densities isosurfaces indicate the formation of 4 covalent bonds. Nonetheless, at this thickness, α-antimonene bonding is still dominant and stronger covalent bonds are formed within α-antimonene bilayers. Interlayer bonding is stronger between the middle bilayers.

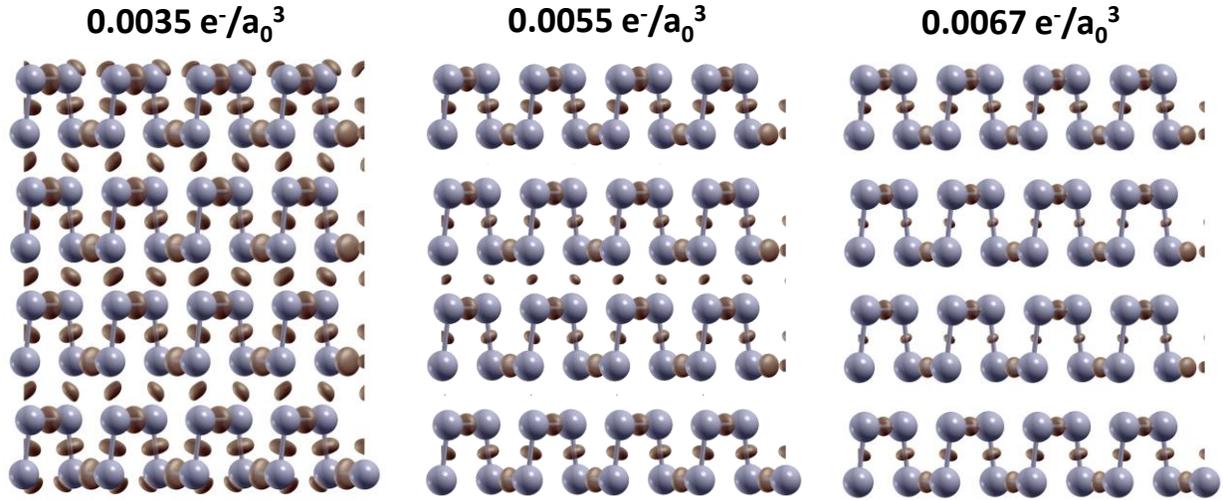

**Figure S12** Charge density minus superposition of atomic charge densities isosurfaces for 4 bilayers thick AA α-antimonene. The 4 bilayers thick film has an α-antimonene structure, but inter-bilayer bonding is observed, especially in the middle of the film.

**Estimation of critical nucleus size and nucleus surface energy**

We can estimate the critical nucleus size $r^*$ and the nucleus surface energy $\gamma$ using a lower limit on the observed nucleation rate of $f = f_0 e^{-\Delta G^*/kT} \approx 0.1\ Hz$, with $\Delta G^* = \pi h \gamma^2 / \Delta G_V$ the free energy of the critical nucleus (cylindrical nucleus), $f_0 = N * 10^{11}\ Hz$, with $N$ the estimated number of nucleation sites (island area divided by nucleus area) and $\Delta G_V = 0.036\ meV/\text{Å}^3$ at a thickness of $h = 45\ \text{Å}$ (8 bilayers). We find $r^* = 18\ \text{Å}$ and $\gamma = 0.64\ meV/\text{Å}^2$.